\documentclass[preprint,12pt]{elsarticle}

\usepackage{amssymb}
\usepackage{amsmath}

\journal{Nuclear Instruments and Methods in Physics Research A}

\begin{document}

\begin{frontmatter}



\title{The T-SDHCAL Hadronic Calorimeter for a Future Higgs factory}

\author[1]{M. Tytgat}
\ead{michael.tytgat@vub.be}
\author[2]{C. Combaret}
\author[2]{C. Devanne}
\author[2]{G. Garillot} 
\author[2]{G. Grenier}
\author[2]{I. Laktineh}
\author[2]{L. Mirabito} 
\author[2]{T. Pasquier}
\author[3]{M.C. Fouz}
\author[4]{H.J. Yang} 
\author[4]{W. Wu}
\author[4]{Y. Tan}
\author[5]{Y. Baek}
\author[5]{D. Kim}
\author[5]{J. Kim}

\affiliation[1]{
    organization={Vrije Universiteit Brussel - Inter-University
    Institute for High Energies},
    postcode={1050},
    city={Brussels},
    country={Belgium}
}
\affiliation[2]{
    organization={Univ. Lyon, Univ. Claude Bernard Lyon 1, CNRS/IP2I Lyon},
    postcode={F-69622},
    city={Villeurbanne},
    country={France}
}
\affiliation[3]{
    organization={CIEMAT, Centro de Investigaciones Energeticas, Medioambientales y Tecnologicas},
    postcode={28040},
    city={Madrid},
    country={Spain}
}
\affiliation[4]{
    organization={Tsung-Dao Lee Institute, Institute of Nuclear and Particle Physics, School of Physics and Astronomy, Shanghai Jiao Tong University, Key Laboratory for Particle Physics, Astrophysics and Cosmology (Ministry of Education), Shanghai Key Laboratory for Particle Physics and Cosmology},
    city={Shanghai},
    postcode={200240},
    country={P.R. China}
}
\affiliation[5]{
    organization={Gangneung-Wonju National University},
    postcode={25457},
    city={Gangneung},
    country={South Korea}
}

\begin{abstract}
The CALICE technological RPC-based SDHCAL prototype that fullfils all the requirements of compactness, hermeticity and power budget of the future lepton accelerator experiments, has been extensively tested and has provided excellent results in terms of the energy resolution and shower separation.
A new phase of R\&D to validate completely the SDHCAL option for the International Linear Detector (ILD) project of the ILC and also the Circular Electron Positron Collider (CEPC and FCCee) has started with the conception and the realization of new prototypes.
The new prototype proposes to exploit the excellent time resolution that can be provided by multi-gap resistive plate chamber 
detectors in order to better build the hadronic showers with the aim to separate close-by ones and also to single out the contribution of delayed neutrons with the purpose to improve on the Particle Flow Algorithm performances and better reconstruct the showers energy.
A new technique to build multi-gap resistive plate chambers has been developed and first results confirm the excellent efficiency of the new detectors. The timing performance is under study using the PETIROC ASIC developed by OMEGA group. 
The progress realized on the different aspects of the new concept will be presented and the future steps will be discussed.
\end{abstract}




\end{frontmatter}



\section{Introduction}
\label{sec1}
Electromagnetic (ECAL) and hadronic (HCAL) calorimeters are crucial components in detectors designed for high-energy physics (HEP) experiments. Their primary role is to measure the energy of charged and neutral particles, as well as to determine the total energy of events. However, these calorimeters also play a significant role in broader event reconstruction, particle identification, and triggering systems.

The ambitious physics goals and challenging experimental conditions anticipated at future colliders necessitate advancements in calorimeter technology to fully harness the potential of these facilities. At future $e^+e^-$ colliders, often called Higgs Factories, achieving unprecedented precision in event reconstruction is a primary objective. Meanwhile, future hadron colliders operating at energies and luminosities significantly beyond those of the High Luminosity Large Hadron Collider (HL-LHC) present unique challenges, particularly in terms of managing the experimental environment.

Incorporating timing information into calorimeter systems offers significant opportunities for both technological enhancements and advanced reconstruction techniques. The ability to do high-precision time measurements in calorimeter systems provides several layers of benefits for event reconstruction, such as (i) timing at cell level, i.e. in highly granular calorimeters, precise timing at the individual cell level aids in the detailed reconstruction of particle showers and enables energy corrections, improving the accuracy of measurements (ii) shower timing, i.e. accurate timing of individual particle showers enhances particle identification and improves the reconstruction of complex objects, such as jets or hadronic decays (iii) object level timing, i.e. leveraging timing information for reconstructed objects facilitates the mitigation of event pile-up, which is a significant challenge at high-luminosity colliders, and aids in the detailed characterization of overlapping events.  When fully exploited, adding such timing capabilities can therefore transform the calorimeter performance, enabling high precision measurements and robust event reconstruction in increasingly complex experimental scenarios. 

This paper summarizes the original Semi-Digital Hadron Calorimeter (SDHCAL) project and describes its extension towards high-time precision calorimetry. 

\section{The Semi-Digital Hadron Calorimeter}
\subsection{Detector design}
The SDHCAL is a stainless steel hadronic sampling calorimeter developed using Glass Resistive Plate Chambers (GRPC) as the sensitive medium, with embedded pad readout electronics. It represents the first technological prototype of a Particle Flow Algorithm (PFA)-oriented calorimeter that was developed within the CALICE Collaboration. The SDHCAL was originally intended and designed for the ILD (International Large Detector)~\cite{ild} of the International Linear Collider (ILC) project~\cite{ilc}, while it is currently also being considered for other future collider options such as the Circular Electron Positron Collider (CEPC)~\cite{cepc} and lepton collider option of the Future Circular Collider (FCC-ee)~\cite{fccee}.  

For the ILD, a unique very compact mechanical structure was devised to avoid dead zones and projective cracks, and minizing the separation between the barrel and endcap part of the detector. This led to a design with variable lengths for the sensitive medium with a maximum size of $0.9\times 3$~m$^2$. This innovative design addresses several challenges, including ensuring GRPC response homogeneity over large areas, creating an active detector with a thickness of only a few millimeters (while integrating low-power consumption embedded electronics), and positioning all services on one side of the detector.

\begin{figure}[ht]
\centering
\includegraphics[width=0.48\textwidth]{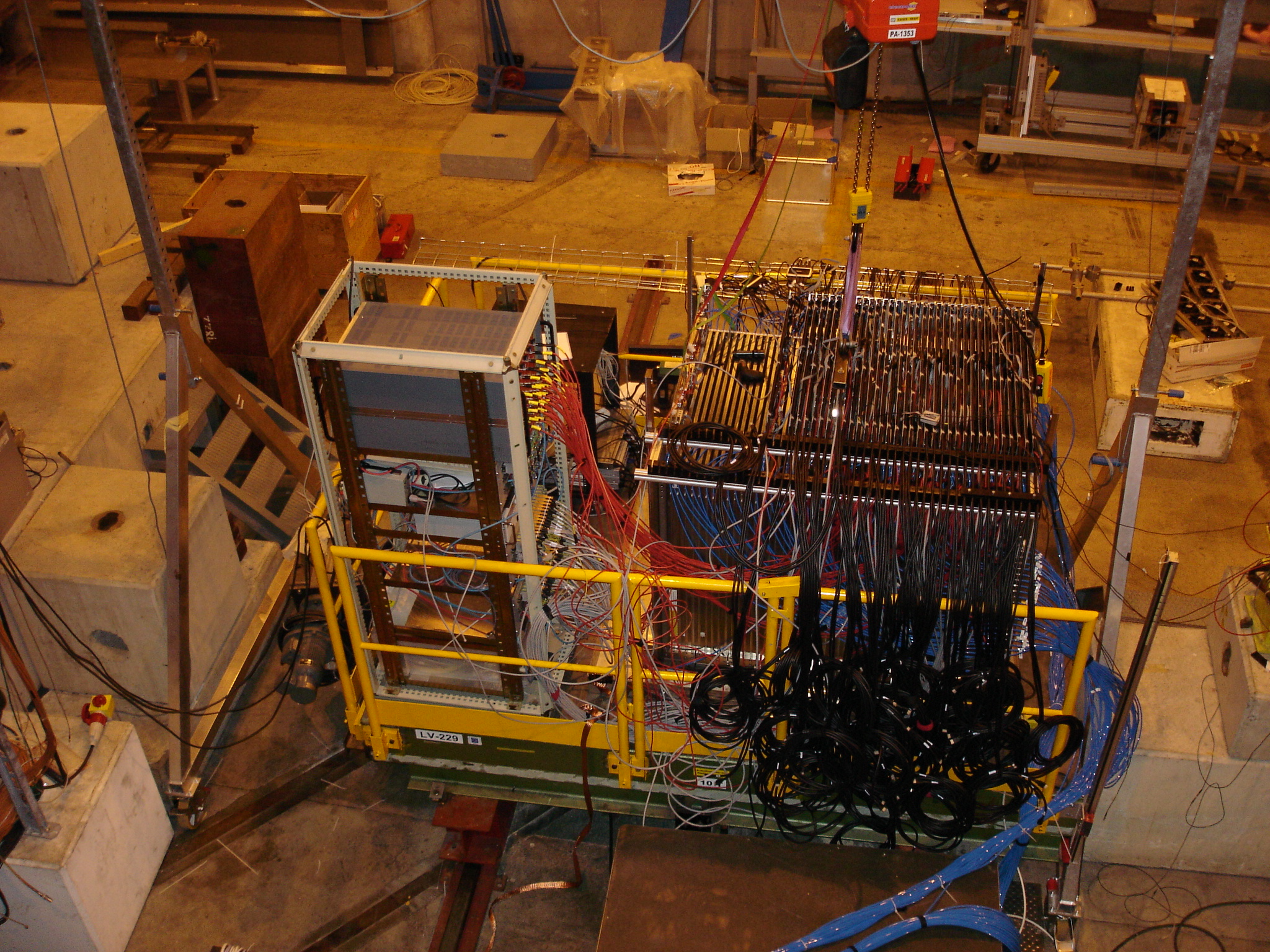}
\includegraphics[width=0.48\textwidth]{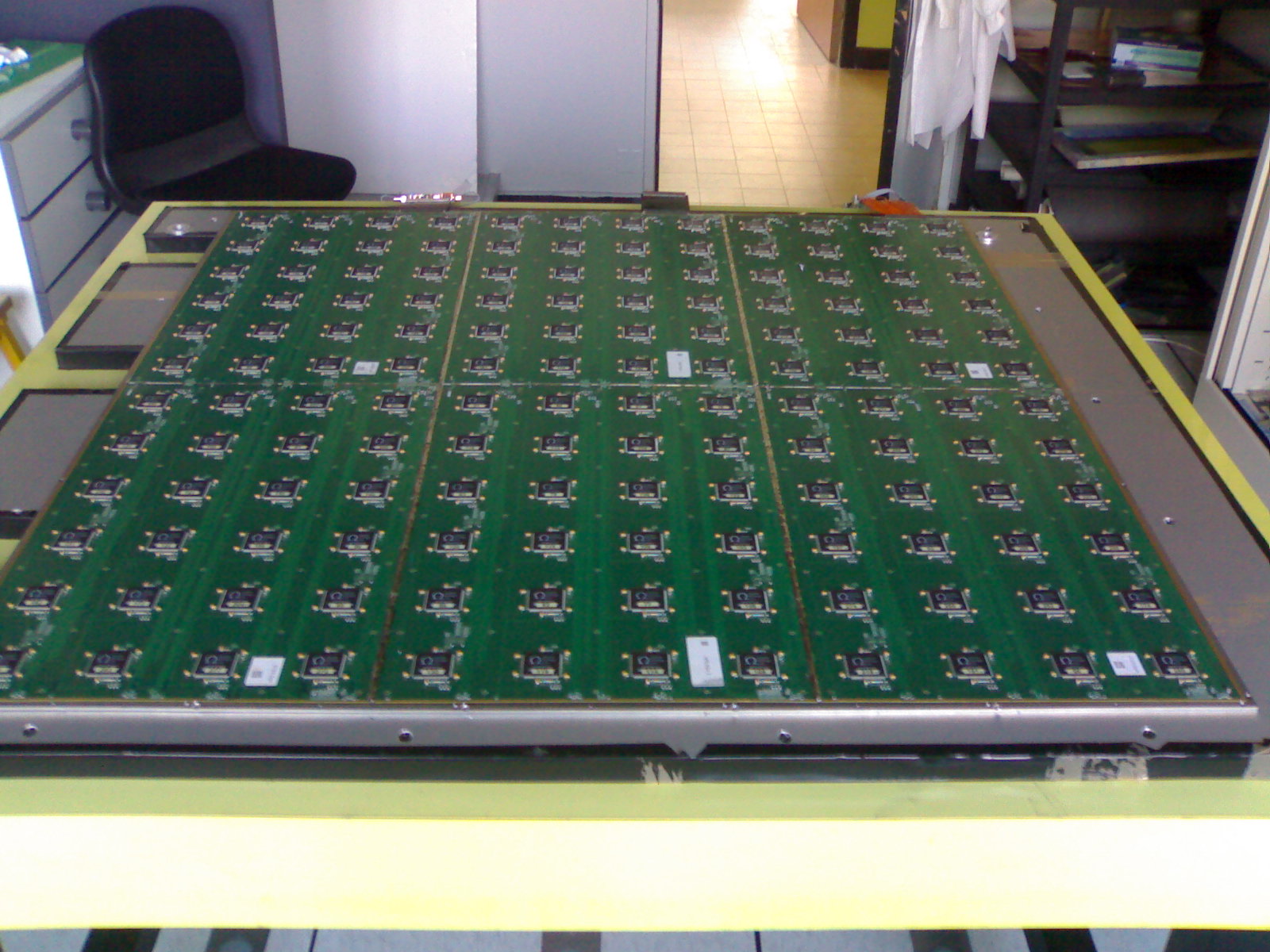}
\caption{The SDHCAL prototype installed at the CERN SPS test beam facility (left), and a 1m$^2$ SDHCAL GRPC readout board containing 144 HARDROC2 ASICs (right), with a grid of 1cm$^2$ readout pads included in the side coupled to the GRPC.}\label{fig:sdhcal_prototype}
\end{figure}

A fully operational SDHCAL prototype of about 1.3~m$^3$, as shown in Fig.~\ref{fig:sdhcal_prototype}, has been conceived as a demonstrator, resolving all these challenges~\cite{sdhcal2015}. It features a stainless steel self-supporting structure capable of accommodating up to 50 GRPC cassettes. Made of two 2.5~mm stainless steel plates, each cassette is 11~mm thick and holds a 6~mm thick GRPC unit as active layer along with its embedded electronics. The GRPCs have a 1~m$^2$ active surface and consist of a single 1.2~mm gas gap made with 0.7~mm and 1.1~mm thick float glass plates with a resistive paint cover. Each GRPC is read out using an array of $96 \times 96$ Copper pads, each with an area of 1~cm$^2$. The 1~m$^2$ custom readout board is constructed by connecting together six smaller, approximately $0.5 \times 0.3$~m$^2$ PCBs called Active Sensor Units (ASU). The full plane is then readout via 3 Detector Interface (DIF) boards each coupled to 2 ASU, and connected to Data Concentrator Cards (DCC) that interface with the data-acquisition system which employs USB and HTML protocols. 

The SDHCAL prototype comprises over 460,000 readout channels, of which less than 1\% are non-functional. The readout electronics employs daisy-chained HARDROC2 chips, each capable of reading an $8\times 8$ array of pads. These ASICs feature a 3-level discriminator, covering a dynamic range from 10~fC to 30~pC, and provide 2-bit, i.e. three threshold levels, readout per pad (hence the name "semi-digital" calorimeter). 
This multi-threshold approach enhances energy estimation at high energies, particularly in dense hadronic showers where individual pads may register signals from tens of particles.

To achieve the target low power consumption of $\leq 10$~$\mu$W per channel, the HARDROC2 chips operate in a power-pulsing mode aligned with the ILC beam time cycle. Each chip records signals for a duration of 1~ms, storing up to 127 threshold crossings in its internal memory. After this period, the chip transmits the stored data and switches off for nearly 200~ms before reactivating for the next data acquisition cycle. Extensive tests have demonstrated that the power-pulsing mode does not compromise detector performance, provided the ASICs are powered on at least 25~$\mu$s before recording events.

\subsection{Results from test beam measurements}

With initial beam tests of small prototype GRPCs starting in 2009, the full-scale SDHCAL prototype has been exposed to a variety of particle beams at the CERN PS and SPS test beam facilities in the period from 2012 to 2022. During these tests, the GRPC units have been operated with a mixture of TetraFluoroEthane (TFE, 93\%), CO$_2$ (5\%) and SF$_6$ (2\%).

The detector response to pion, electron and muons beam has been verified, yielding preliminary methods for particle identification based on the observed 3D shower shapes and features. An excellent energy reconstruction capability has been obtained~\cite{sdhcal2016, sdhcal2022} for beam energies ranging from a few GeV to about 80~GeV, which is the typical particle energy range that calorimeters are expected to see at future colliders.  Especially at higher particle energies, the multi-threshold (semi-digital) feature of the SDHCAL was shown to yield superior energy resolution compared to a purely binary readout mode. The high granularity of the SDHCAL actually allowed the reconstruction via a 3D Hough transform technique of track segments inside particle showers~\cite{CALICE:2017nol}. These track segments allow in-situ layer efficiency and multiplicity studies, and can improve the PFA algorithm and particle energy reconstruction. Furthermore, more refined particle identification studies have been done based on multi-variate techniques, where boosted decision trees taking shower features as input yielded e.g. electron and muon rejection factor above 99\%~\cite{CALICE:2020kww}. 

\subsection{Further SDHCAL developments}

The ILD (International Large Detector) design requires the development of Glass Resistive Plate Chambers (GRPCs) with lengths of up to 3~m. To meet this requirement, prototypes of 2~m GRPCs have been constructed so far, featuring a scalable gas distribution system integrated within the chamber. This system improves the gas flow by ensuring a homogeneous velocity profile, thereby optimizing the gas distribution across the chamber.

To accommodate GRPCs of sizes up to $1\times 3$~m$^2$, a new mechanical structure has been designed and built. This structure incorporates plates processed through roller leveling techniques to achieve precise flatness and assembled using electron beam welding. These advanced manufacturing methods ensure that the aplanarity of the GRPC cassette holding slots remains below 1~mm, as traditional screw and bolt assembly techniques were found insufficient for this level of precision.

In parallel, the electronics system has been entirely redesigned to meet the demands of the ILD. The new PCBs, which host 1~cm$^2$ pads, measure $1\times0.3$~m$^2$ and can be chained together along their longer sides. They are equipped with the upgraded HARDROC3 chip, which features an extended dynamic range of up to 50 fC. Additionally, the clock distribution, slow control, and fast control systems have been enhanced to improve communication efficiency across long chains of ASICs and PCBs. The clock distribution employs the TTC protocol, while the slow and fast controls use the I2C protocol, ensuring reliable and rapid communication.

To support the redesigned electronics, a new Detector Interface Board has been developed. This board is capable of directly communicating with up to 432 ASICs through I2C links, providing sufficient capacity to handle the maximum GRPC size envisioned for the ILD. Together, these innovations in mechanical structure, gas distribution, and electronics enable to meet the requirements of the ILD. 

\begin{figure}[ht]
\includegraphics[width=0.48\textwidth]{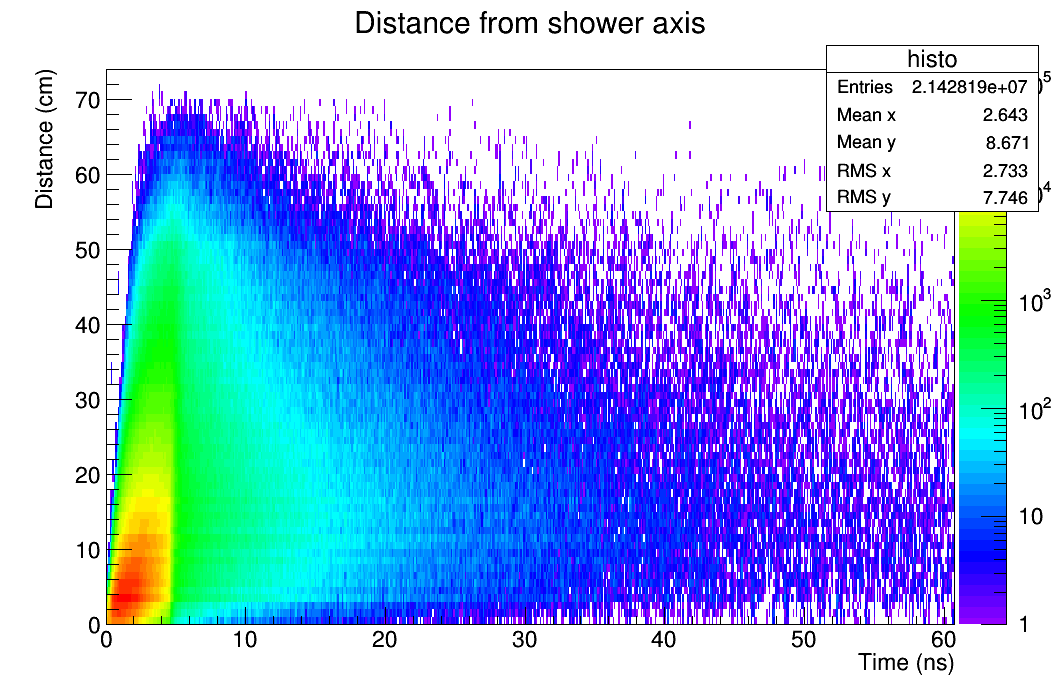}
\includegraphics[width=0.48\textwidth]{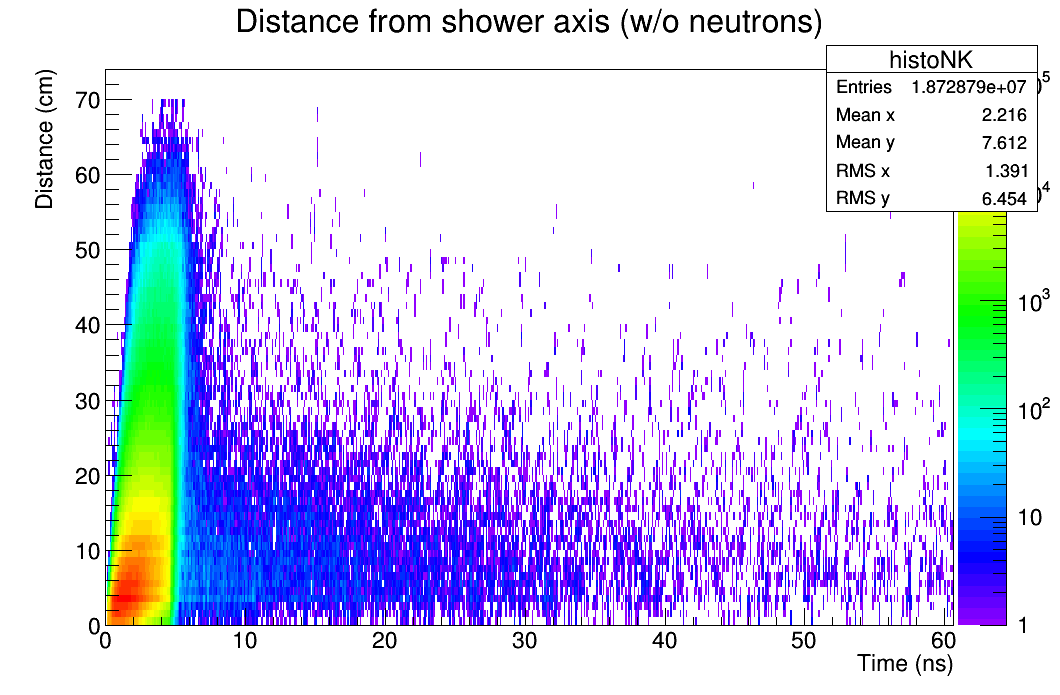}
\caption{Simulated distribution of hits in the SDHCAL as function of their distance from the shower axis and the hit time, for all hits (left) and for all hits except those induced by neutrons (right).} 
\label{fig:tsdhcal_neutrons}
\end{figure}

\begin{figure}[ht]
\includegraphics[width=\textwidth]{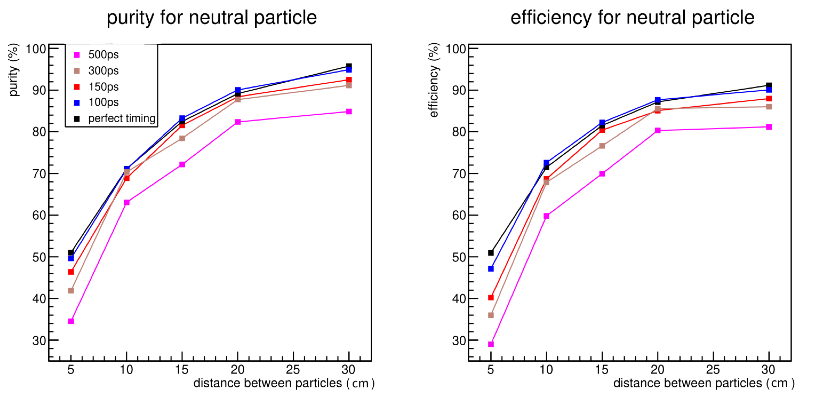}
\caption{Simulated purity (left) and reconstruction efficiency (right) for a 10~GeV neutral particle in the presence of a nearby 30~GeV charged particle in the SDHCAL.} 
\label{fig:tsdhcal_showersep}
\end{figure}

\section{The Timing Semi-Digital Hadron Calorimeter}

As mentioned above, implementing precision timing in calorimeters can bring many general benefits for future collider experiments. As an example, hadronic showers are known to exhibit a complex temporal structure, due to e.g. delayed components linked to neutron-induced interactions as shown in Fig.~\ref{fig:tsdhcal_neutrons}, where the neutron-induced segment of the shower appears more diffuse and spatially extended compared to the electromagnetic and relativistic hadronic components. These 
neutron-induced hits not only cause event-by-event fluctuations, which has a negative impact on the accuracy of the shower energy reconstruction, their spatial distribution also increases the overlap between closely positioned showers, reducing the ability to distinguish them effectively. As such, incorporating timing at the cell level could significantly enhance the spatial reconstruction of these showers. A time resolution in the range of a few hundred picoseconds to one nanosecond would allow for a sharper definition of the shower core, which can improve particle separation capabilities within the calorimeter and enhance the 
track-cluster matching in PFA~\cite{chekanov2022precisiontimingcolliderexperimentbasedcalorimetry}. 
The issue with separation of hadronic showers in high-density environments is also illustrated in Fig.~\ref{fig:tsdhcal_showersep}, where the example is given for the overlap between a 10~GeV neutral and 30~GeV charged particle in the SDHCAL.
Current shower reconstruction algorithms rely on spatial positioning to associate nearby hits, grouping them into a single shower even if they were induced at different times.
Without time information, distinguishing two hadrons entering the SDHCAL within 10 cm of each other becomes very challenging, with efficiencies and purities dropping significantly. A substantial improvement could be achieved with a time resolution that surpasses the limit set by the causality relationship between adjacent hits, which is a time resolution comparable to the calorimeter's spatial granularity multiplied by the speed of light, i.e. in the range of several tens to few hundreds of picoseconds.

The ongoing R\&D within this project includes the aim to replace the single-gap GRPCs of the existing SDHCAL prototype with multigap glass resistive plate chambers (MGRPC). MGRPC technology offers significantly enhanced time resolution, achieving values on the order of 100 ps or lower, compared to the sub-nanosecond performance of single-gap GRPCs. To profit of these improved timing detectors, the HARDROC chip will in a first step be substituted by a version of the OMEGA PETIROC ASIC~\cite{petiroc}. The latter was already successfully tested in the context of the Phase-2 RPC upgrade of the CMS experiment~\cite{cmsrpcupgrade}. The improved detector timing capability, combined with the PETIROC chip's ability to measure both energy and time with a resolution better than 20~ps, enables a total time resolution for individual crossing pads of approximately 150~ps. This would pave the way for transforming the existing SDHCAL into a 5D Timing-SDHCAL (T-SDHCAL) prototype, incorporating energy ($E$), spatial ($x$,$y$,$z$), and temporal ($t$) dimensions. 

\begin{figure}[ht]
\centering
\includegraphics[width=0.35\textwidth]{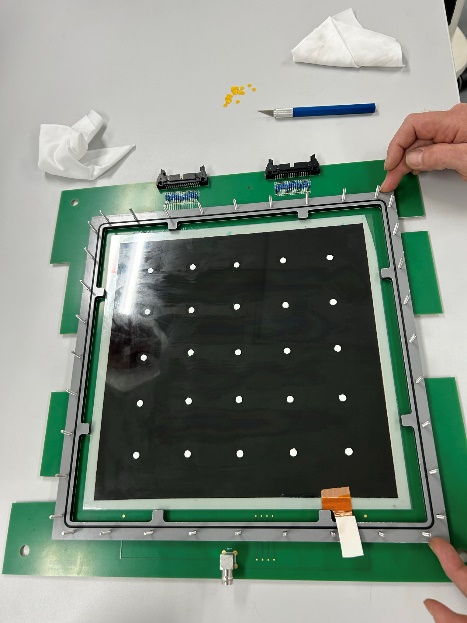}
\includegraphics[width=0.63\textwidth]{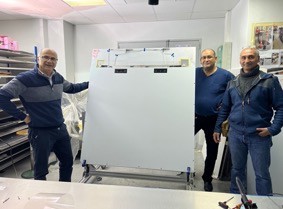}
\caption{A small, $30\times 30$~cm$^2$ MGRPC with 6 250~$\mu$m gaps (left), and a 1~m$^2$ 4-gap MGRPC prototype (right), both assembled using the new circular spacer method.}
\label{fig:MGRPC}
\end{figure}

\subsection{MGRPCs for the T-SDHCAL}

Within the T-SDHCAL project, various different versions of MGRPCs in terms of detector area, assembly procedure, gas gap sizes, single and double stack structures, and readout boards are being produced and tested. Using the regular assembly method with fishing wire to define the gas gap sizes, efficiencies above 92\% have been achieved for 4-gap, 1~m$^2$ area chambers. Given the issues that have been observed with increased dark current and noise appearing near the fishing wires ~\cite{PETRIS2024169584}, a new simplified detector assembly procedure has been implemented in which the wires are replaced by mylar spacers. With this new method (see Fig.~\ref{fig:MGRPC}), the gas gaps are constructed using circular pieces cut out of a mylar foil, each spacer with a diameter of a few mm and a thickness equal to the desired gas gap size. The spacers are attached using double-sided adhesive tape to the glass plates, at evenly distributed locations across the surface. This new procedure simplifies the detector design, e.g. avoiding the need for any structure to fix and tension the fishing wires. Several MGRPC prototypes have been produced in this way, from $30 \times 30$~cm$^2$ up to 1~m$^2$ surface area, where detector efficiencies exceeding 95\% have been achieved. 

As the new MGRPC units have to fit inside the mechanical cassettes of the existing SDHCAL prototype, the next step in the project is to optimize the MGRPC gap configuration in terms of number and size of the gaps, to achieve optimal time resolution for a maximum detector thickness.

\begin{figure}[ht]
\centering
\includegraphics[width=0.37\textwidth]{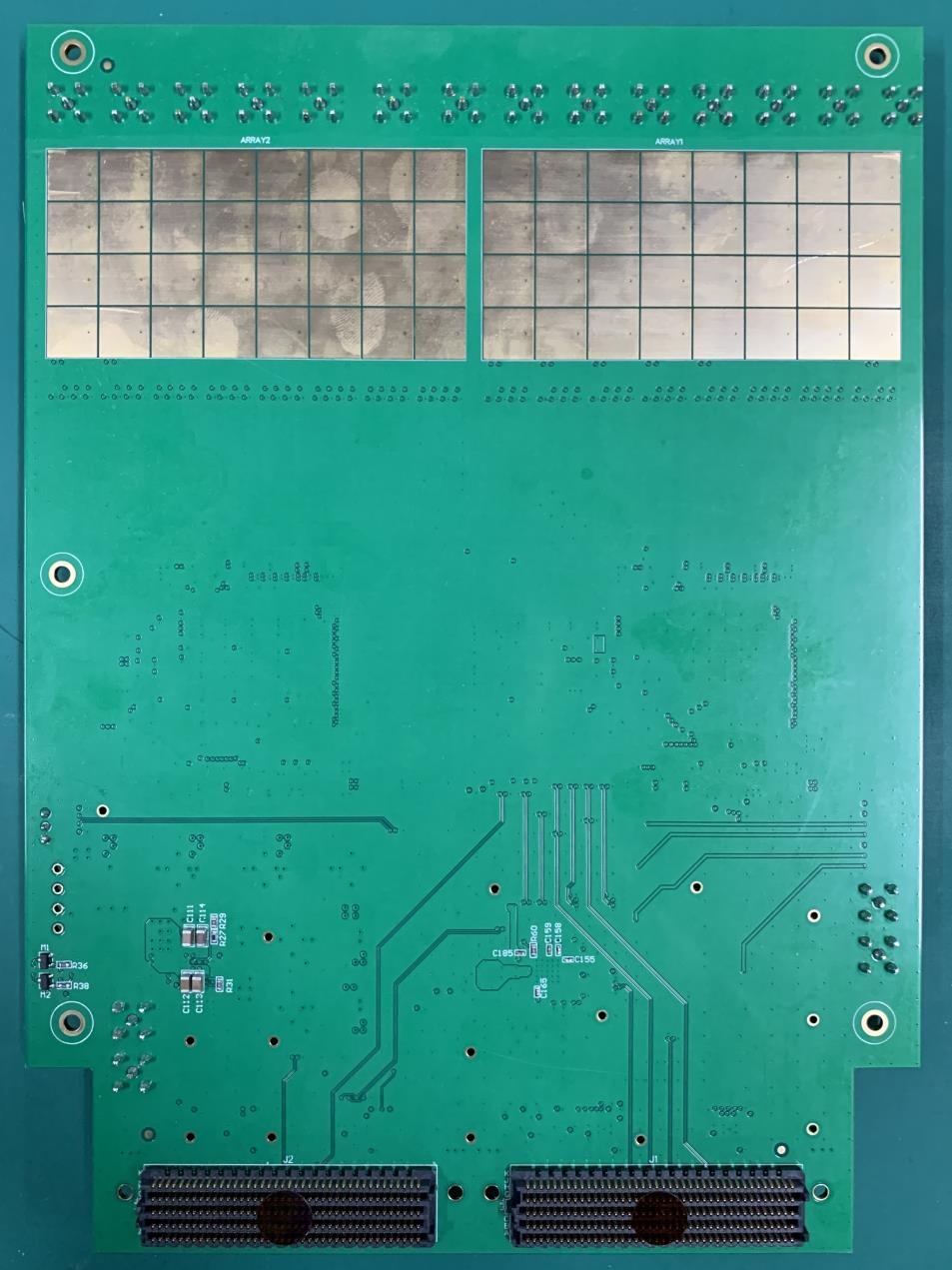}
\includegraphics[width=0.37\textwidth]{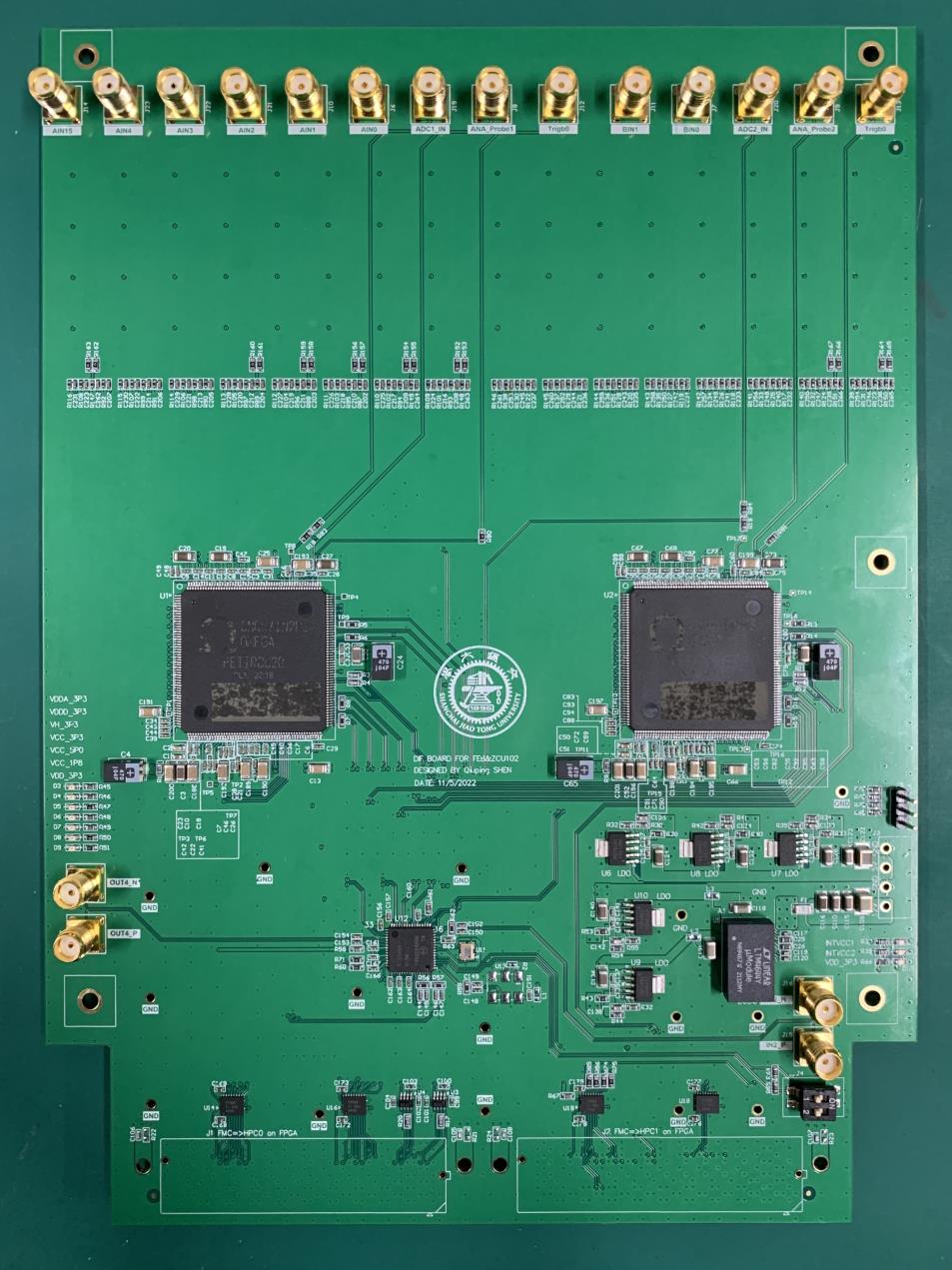}
\caption{Second version of the PETIROC-based prototype readout PCB with 2 ASICs and 64 readout pads.}\label{fig:PCBv2}
\end{figure}

\subsection{T-SDHCAL front-end and data-acquisition}

The present baseline solution for the MGRPC readout with improved timing uses the 32-channel OMEGA PETIROC ASIC, which has an embedded pre-amp, on-chip QDC and TDC functionality, and a time resolution below 50~ps. Several versions of small prototype PCBs with 1~cm$^2$ pickup pads have been conceived to readout initial MGRPC chambers. The second version of the prototype board as shown in Fig.~\ref{fig:PCBv2} houses 2 PETIROC ASICs, has 64-channel input pads and uses SMA connectors to inject test signals. Injection tests with a function generator to evaluate the timing performance yielded an intrinsic time resolution of 36~ps. The current idea is to cover a 1~m$^2$ chamber with 4 separate boards, each housing 16 PETIROCs (i.e. 512 channels) and linked together with tiny connectors. The readout pads would be 1.5~cm$^2$ such that a 10-layer structure for signal routing would suffice. 

\begin{figure}[ht]
\centering
\includegraphics[width=0.65\textwidth]{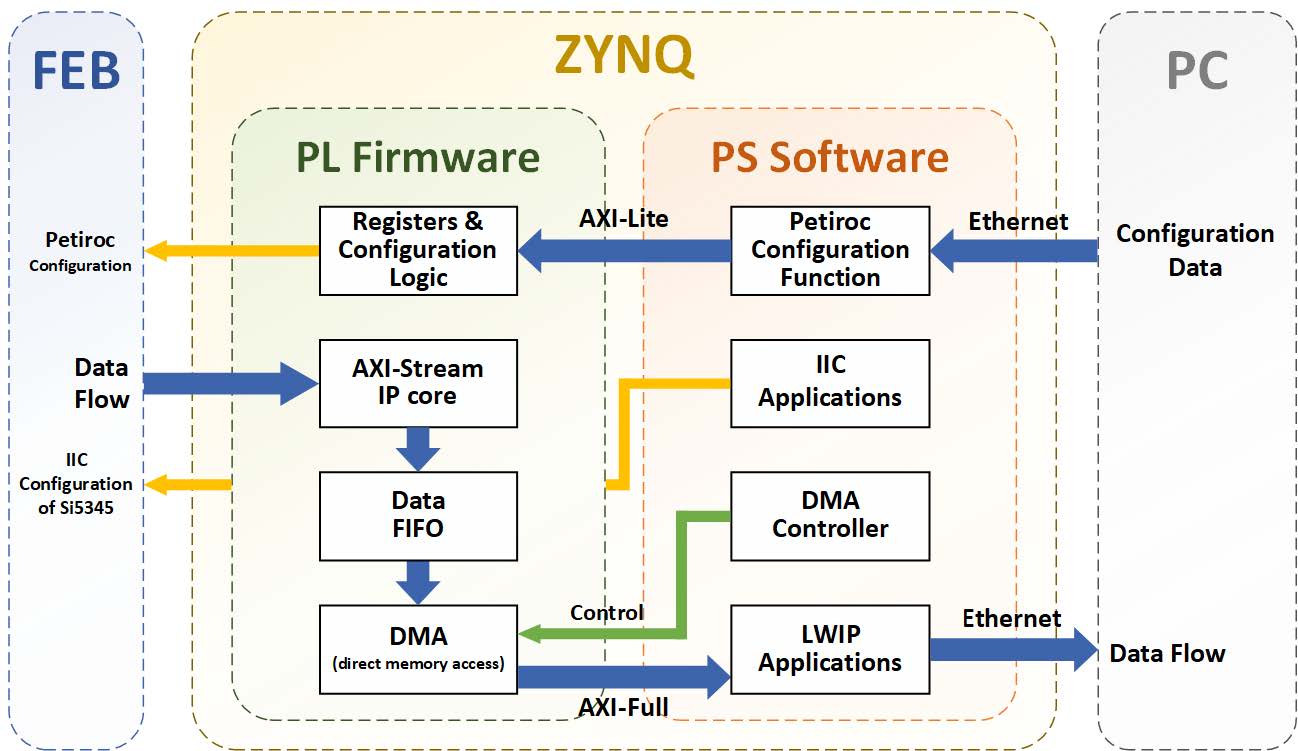}
\caption{Layout of the data-acquisition under development for the MGRPC-based TSDHCAL.}
\label{fig:daq}
\end{figure}

Along with the new front-end, a corresponding new data-acquisition system is being developed based on the commercial ZCU102 FPGA evaluation board with the Zynq UltraScale+ system-on-chip. As depicted in Fig.~\ref{fig:daq}, it features a processing system with light-weight IP data transmission and a programmable logic for the system configuration. The data-acquisition software is being developed in Python with a Qt5-based graphical interface.

\section{Summary and future prospects}

The T-SDHCAL project aims to develop a new generation calorimeter candidate for a future Higgs factory detector. Starting from the existing SDHCAL prototype, replacing the single-gap GRPCs with high-time resolution MGRPCs should yield a 5D calorimeter prototype with a timing precision on the order of 100~ps. Various MGRPC configurations are currently being tested, where also a new assembly technique using small mylar spacers to define the gas gaps has been successfully implemented for chambers areas up to 1~m$^2$. 
To exploit the improved MGRPC time resolution new front-end readout boards based on the PETIROC ASIC are being designed and tested. 

As the SDHCAL was originally developed for the ILC, which has relatively repetition rate and allows for a power pulsed readout, modifications will be needed to adapt the SDHCAL for other future collider options. This means among others that the readout system will need to be replaced to enable continuous detector readout. As consequence, the power consumption is expected to increase by a factor of about 100-200 such that an active cooling system will be essential. Simulation studies to develop a water based cooling system with Copper tubes in contact with the ASICs, keeping in mind the SDHCAL mechanical structure as well as the power consumption of the new ASICs, will be carried out.   

As an alternative for the PETIROC-based readout for the MGRPC, to achieve even better time resolution, the 64-channel LIROC ASIC~\cite{liroc} will be considered, which has a FWHM time resolution below 20~ps. As this ASIC does not have an internal TDC, we would combine it with the 64-channel PicoTDC that offers a  3~ps or 12~ps binning with very low jitter~\cite{picotdc}.

Furthermore, as the new MGRPCs will have to cope with high rates of a few thousands of particles per second, the use of alternative electrode materials with low resistivity will be explored to improve the calorimeter rate capability. Finally, also the use of new, eco-friendly RPC gas mixtures will be considered (see for example ~\cite{RPCECOGasGIF:2024ilv, rpcecogas_cbmtof}), where special attention will be given to the timing performance of the MGRPC. 

\bibliographystyle{elsarticle-harv} 
\bibliography{tsdhcal}



\end{document}